# Graphene-on-Sapphire and Graphene-on-Glass: Raman Spectroscopy Study

Irene Calizo

Nano-Device Laboratory, Department of Electrical Engineering
University of California – Riverside, Riverside, California 92521

W. Bao, F. Miao and C.N. Lau
Department of Physics and Astronomy
University of California – Riverside, Riverside, California 92521

Alexander A. Balandin[*]
Nano-Device Laboratory, Department of Electrical Engineering
University of California – Riverside, Riverside, California 92521

## Abstract

The room-temperature Raman signatures from graphene layers on sapphire and glass substrates were compared with those from graphene on GaAs substrate and on the standard Si/SiO$_2$ substrate, which served as a reference. It was found that while $G$ peak of graphene on Si/SiO$_2$ and GaAs is positioned at 1580 cm$^{-1}$ it is down-shifted by ~5 cm$^{-1}$ for graphene-on-sapphire (GOS) and, in many cases, splits into doublets for graphene-on-glass (GOG) with the central frequency around 1580 cm$^{-1}$. The obtained results are important for graphene characterization and its proposed graphene applications in electronic devices.

[*] Corresponding author; electronic address (A.A. Balandin): balandin@ee.ucr.edu ; web: http://www.ndl.ee.ucr.edu/

Graphene has attracted major attention from the physics and device research communities [1-6]. In addition to its unusual physical properties it also shows a promise as a material for the electronic applications. Geim and Novoselov [7] suggested that a band gap can be induced in the bi-layer graphene (BLG) and engineered in the single-layer graphene (SLG) by the spatial confinement or lateral superlattice-type potential. The extremely high room temperature carrier mobility (up to 15,000 $cm^2V^{-1}s^{-1}$ [1-3]) represents an advantage over Si, making graphene a candidate for applications in the circuits beyond the conventional complementary metal-oxide-semiconductor technology. Raman spectroscopy has been successfully utilized as a convenient technique for identifying and counting graphene layers [8-13]. Specifically, it was shown [8] that the evolution of the *2D* band Raman signatures with the addition of each extra layer of graphene can be used to accurately count the number of layers. We have recently reported the temperature dependence of the *G* peak and *2D* band in graphene on Si/$SiO_2$ substrates [10-11]. The extracted values of the temperature coefficients $\chi_G$ for the *G* peak in the spectra of SLG and BLG are $-(1.6\pm0.2)\times10^{-2}cm^{-1}/K$ and $-(1.5\pm0.06)\times10^{-2}cm^{-1}/K$, respectively [11].

Most, if not all, Raman spectroscopy studies of graphene reported to date were limited to either graphene layers on Si/$SiO_2$ substrates [8-13] with a very carefully selected thickness of the $SiO_2$ layer, or to tiny dispersed flat carbon clusters, which have also been referred to as graphene [14-15]. The reason for choosing a specific substrate for the mechanically exfoliated graphene is the observation that it becomes visible in an optical microscope when placed on top of Si wafer with 300-nm thick oxide layer [1-2]. Thus, it is easier to carry out Raman spectroscopy of graphene layers on the standard Si/$SiO_2$ (300 nm) substrates because one can pin-point the exact location of a graphene sample (which typically has the lateral dimensions of few micrometers) and carry out an initial identification of the number of layers under the optical microscope. Future studies of graphene's unique properties and its application as an electronic material call for graphene integration with a variety of different materials and substrates. However, presently very little is known about the visibility or property of graphene on substrates other than Si/$SiO_2$, and there is no confirmed experimental tool for determining the number of layers in few-layer graphites on these substrates. Thus, it is useful to expand Raman spectroscopy as a nanometrology tool for graphene and graphene-based devices to various substrates. Another important motivation for the study of the substrate influence on graphene Raman spectrum is a fundamental question of the role played by the

graphene – substrate interface. The measurements of Raman spectra from graphene on different substrates can shed light on the strength of the graphene – substrate coupling.

In this letter we report the room-temperature spectroscopic Raman microscopy of the single-layer and few-layer graphene (FLG) deposited on different substrates. SLG and FLG were obtained by micromechanical cleavage of bulk graphite using the process outline in Refs. [1-2]. An identical procedure was used to place graphene layers on a reference Si/SiO$_2$ (300 nm) substrate and on a set of distinctively different substrates, which included n-type (100) GaAs wafer, A-plane (11-20) sapphire (Al$_2$O$_3$) and glass substrates. The number of layers was determined from the visual inspection of graphene on Si/SiO$_2$ (300 nm), atomic force microscopy (AFM) and analysis of the *2D* band features using the approach outlined in Ref. [8]. For GaAs substrate, we only succeeded in transferring five-layer graphene as confirmed by AFM and Raman spectrum of the *2D* band. The AFM inspection of graphene on sapphire and glass substrates revealed spots with thickness <2 nm, indicating the presence of less than 4 layers. Figs. 1 (a) and (b) show AFM images of graphene on Si/SiO$_2$ and glass substrates, respectively. The following Raman analysis allowed us to conclude that the transferred graphene samples on sapphire and glass are most likely SLG. In order to provide additional verification for the number of layers and graphene quality we carried out transport studies for some of the samples on the Si/SiO$_2$ substrate by attaching the electrodes using the standard nanofabrication techniques, which we described elsewhere [6, 10]. The electrical measurements were performed at low temperature in a sorption pumped $^3$He refrigerator. The extracted high values of the carrier mobility (~8,000 – 15,000 cm$^2$/Vs) and the anomalous "half-integer" plateau, which is a unique signature of the band structure of graphene, attested to the high quality of our samples.

The Raman microscopy was carried out using the Renishaw instrument under 488-nm excitation at low power level to avoid the laser heating effects [16]. A Leica optical microscope with a 50x objective was used to collect the backscattered light from the graphene samples. The Rayleigh light was rejected by the holographic notch filter with a 160 cm$^{-1}$ cut off frequency. Since it was important to separate the effect of the substrate from spatial variations in the graphene properties, we took 10-20 spectra in different location for each of the examined samples. A special care has been taken to make sure that all locations for the Raman scans are selected within the sample region with the same number of layers. Fig. 2 (a) presents a close-up of *2D* bands for graphene as the number of layers increases from one to five. The observed features are consistent with the previously reported data [8, 10-11]. After taking Raman spectra from graphene layers on the standard substrate we investigated graphene placed on GaAs, sapphire and glass substrates. The adherence of SLG and FLG to different substrates was similar. To avoid the fabrication damage and charge transfer no contacts were fabricated on the samples subjected to detail Raman study.

Fig. 2 (b) shows a typical spectrum of FLG on n-type GaAs substrate. Two pronounced features in the spectrum are the *G* peak at 1580 cm$^{-1}$ and the *2D* band at ~2736 cm$^{-1}$. The decomposition and analysis of the *2D* band features confirm that the number of layers is five. The measured spectrum features, e.g. *G*-peak position and shape and *2D*-band shape, are very similar to those observed for FLG on the standard Si/SiO$_2$ (300 nm) substrate. Three curves in Fig. 2 (b) correspond to the spectra taken from three different locations. Since there is virtually no variations in the spectra one can conclude that the sample is uniform and the measured results are reproducible. *G* peak recorded for graphene on GaAs substrate is essentially in the same location and of the same shape as the one measured by us [10-11] and others [8-9] for graphene layers on Si/SiO$_2$ (300 nm).

The spectra measured for graphene on the glass and sapphire substrates were much noisier than those for graphene on Si/SiO$_2$ (300 nm) or GaAs substrates. Specifically, the spectra from graphene on a glass substrate manifested a large number of peaks attributed to the amorphous nature of the substrate, which resulted in many local vibrational modes. At the same time, it was always possible to identify *G* peak and *2D* band. Fig. 3 (a) and (b) present a close-up of *G* peak for a single-layer graphene-on-sapphire (GOS) and graphene-on-glass (GOG), respectively. One can see in Fig. 3 (a) that *G* peak in GOS spectra is red-shifted from its position in the spectra from SLG on a standard substrate by ~5 cm$^{-1}$. This shift is observed for all locations; a small spot-to-spot variation in the peak position of about ~1 cm$^{-1}$ is equal to the spectral resolution of the instrument. An unusual feature in the spectra from GOG in Fig. 3 (b) is a splitting of *G* peak into an asymmetric doublet for approximately half of the examined locations. When the *G* peak is not split, it is located at 1579 cm$^{-1}$, which is consistent with its position in graphene on the standard substrate. In the spectra where *G* peak is split, its central frequency is ~1580 cm$^{-1}$. Thus, the *G*-peak position in GOG spectra is close to the one in SLG spectra on the standard Si/SiO$_2$ (300 nm) substrate. The *G*-peak splitting in Raman spectra from some locations on GOG can be attributed to presence of the randomly distributed impurities or surface charges. The localized vibrational modes of the impurities can interact with the extended phonon modes of graphene leading to the observed splitting. The *G*-peak positions and their full width at half maximum (FWHM) for different substrates are summarized in Table I. One can see that FWHM for *G* feature from GOG is the largest. The latter is likely related to the amorphous nature of the glass substrate and inhomogeneous properties of graphene layers on a given substrate.

Table I: Raman *G* Peak Position for Graphene Layers on Different Substrates

| Substrate | G Peak Position (cm$^{-1}$) | G Peak FWHM (cm$^{-1}$) |
| --- | --- | --- |
| Si/SiO$_2$ | 1580 | 15 |
| GaAs | 1580 | 15 |
| Sapphire | 1575 | 20 |
| Glass | 1580[*] | 35 |

[*]This value corresponds to the middle frequency for a doublet if *G* peak is split.

The relatively weak dependence of *G* band on the substrate can be explained by that fact that it is made up of the long-wavelength optical phonons of particular symmetry. The *G*-band optical phonons in graphene represent the in-plane vibrations since the $E_{2g}$ symmetry of this band restricts the atomic motion to the plane of the carbon atoms [17]. According to the first-principle calculations, the out-of-plane vibrations in graphene are not coupled to the in-plane motion [18]. The dependence is stronger for graphene on the A-plane sapphire substrate, where we observed consistent ~5 cm$^{-1}$ shift of *G* mode. The latter can be related to the specifics of the carbon – sapphire binding similar to the phenomenon reported in Ref. [19]. Han et al. [19] observed formation of the highly aligned single-wall carbon nanotube (SW-CNT) arrays on A-plane and R-plane sapphire substrates with negligible miscut, i.e., without apparent involvement of the step edges. Such spontaneous self-orientation was not observed for other types of the substrates. From their AFM studies the authors concluded that strong CNT – sapphire substrate interaction plays a major role in the CNT alignment. Similar interaction forces may lead to the *G*-mode position change in our GOS samples. Another possibility is a presence of the surfaces charges, which lead to the changes in the graphene lattice parameter with the corresponding peak shift.

*Acknowledgements*

AAB acknowledges support from the Focus Center Research Program (FCRP) - Center on Functional Engineered Nano Architectonics (FENA).

**References**

[1] K. S. Novoselov, A. K. Geim, S. V. Morozov, D. Jiang, Y. Zhang, S. V. Dubonos, I. V. Grigorieva, , and A. A. Firsov, Science **306**, 666 (2004).

[2] K. S. Novoselov, A. K. Geim, S. V. Morozov, D. Jiang, M. I. Katsnelson, I. V. Grigorieva, S. V. Dubonos, and A. A. Firsov, Nature **438**, 197 (2005).

[3] Y. B. Zhang, Y. W. Tan, H. L. Stormer, and P. Kim, Nature **438**, 201 (2005).

[4] C.L. Kane and E. J. Mele, Phys. Rev. Lett. **95**, 226801 (2005).

[5] D. V. Khveshchenko, Phys. Rev. B **74**, 161402 (2006).

[6] F. Miao, S. Wijeratne, Y. Zhang, U.C. Coskun, W. Bao, C.N. Lau, Science, **317**, 1530 (2007).

[7] A. K. Geim and K.S. Novoselov, Nature Materials **6**, 183 (2007).

[8] A. C. Ferrari, J. C. Meyer, V. Scardaci, C. Casiraghi, M. Lazzeri, F. Mauri, S. Piscanec, D. Jiang, K. S. Novoselov, S. Roth, and A. K. Geim, Phys. Rev. Lett. **97**, 187401 (2006).

[9] A. Gupta, G. Chen, P. Joshi, S. Tadigadapa, and P. C. Eklund, Nano Letters **6**, 2667 (2006).

[10] I. Calizo, F. Miao, W. Bao and C.N. Lau and A.A. Balandin, Appl. Phys. Lett., 91, 071913 (2007).

[11] I. Calizo, A.A. Balandin, W. Bao, F. Miao and C.N. Lau, Nano Letters, 7, 2645 (2007).

[12] D. Graf, F. Molitor, K. Ensslin, C. Stampfer, A. Jungen, C. Hierold, and L. Wirtz, Nano Letters **7**, 238 (2007).

[13] A. N. Sidorov, M. M. Yazdanpanah, R. Jalilian, P. J. Ouseph, R. W. Cohn, and G. U. Sumanasekera, Nanotechnology **18**, 135301 (2007).

[14] J.F. Cardenas, Chem. Phys. Lett., 430, 367 (2006).

[15] C. Holzl, H. Kuzmany, M. Hulman, J. Wu, K. Mullen, E. Boroviak-Palen, R.J. Kalenczuk and A. Kukovecz, phys. stat. sol. (b), 243, 3142 (2006).

[16] K. A. Alim, V. A. Fonoberov, M. Shamsa, and A. A. Balandin, J. Appl. Phys. **97,** 024313 (2005); K. A. Alim, V. A. Fonoberov, and A. A. Balandin, Appl. Phys. Lett. **86,** 053103 (2005).

[17] Tuinstra, F.; Koenig, J.L. J. Chem. Phys. **53**, 1126 (1970).

[18] Falkovsky, L.A. cond-mat/0702409.

[19] S. Han, X. Liu and C. Zhou, J. Am. Chem. Soc., 127, 5294 (2005).

I. Calizo, W. Bao, F. Miao, C.N. Lau and A.A. Balandin, "Graphene-on-Sapphire and Graphene-on-Glass" 2007

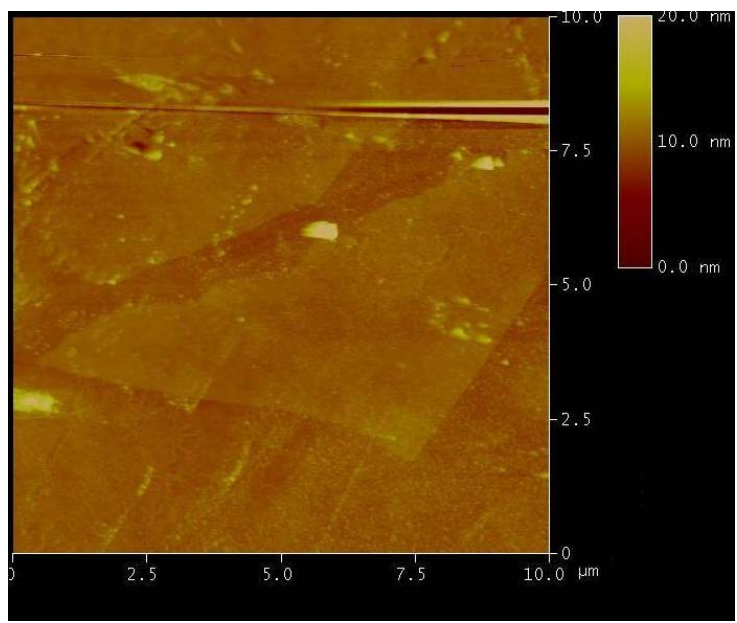

Figure 1: Atomic-force microscopy image of graphene layers on a glass substrate.

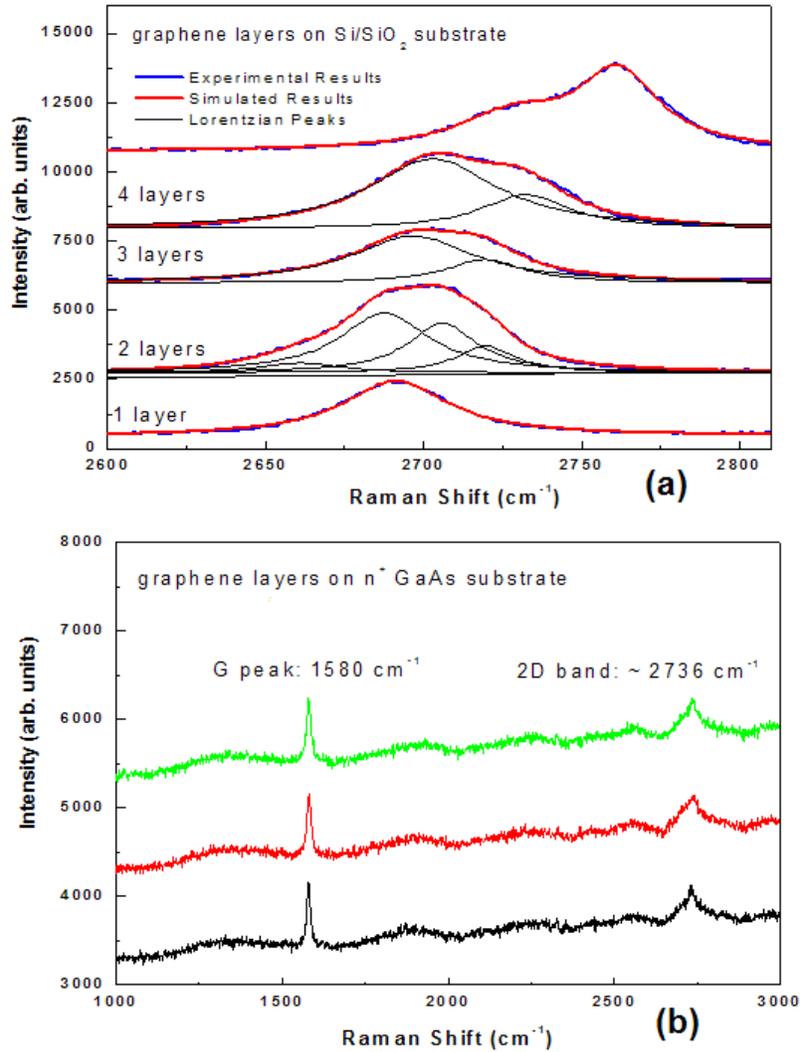

Figure 2: (a) Raman spectrum of *2D* band of graphene on Si/SiO$_2$ substrates as a number of layers changes from one to five. The analysis of the *2D* band was used to verify the number of graphene layers. (b) Raman spectrum of graphene layers on GaAs substrate. Three spectra are taken from different locations on the sample to demonstrate reproducibility and sample uniformity.

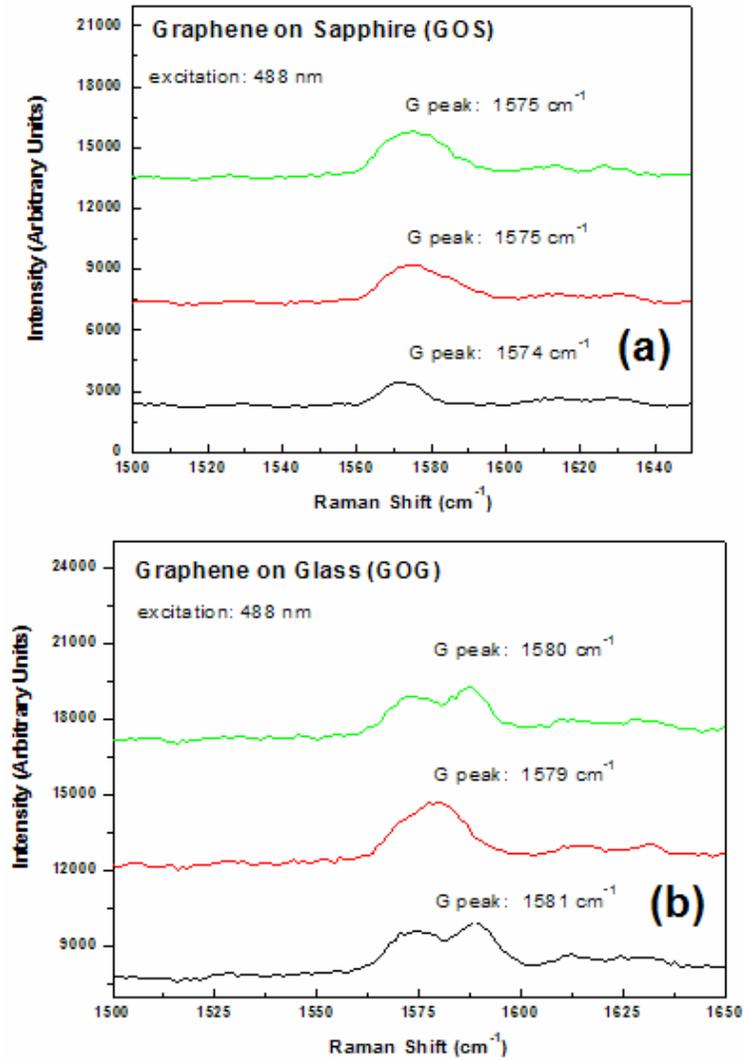

Figure 3: (a) Raman spectra of graphene-on-sapphire (GOS) and (b) Raman spectra of graphene-on-glass. In both cases the *G*-peak region is shown. Three spectra for each substrate are taken from different locations.